\documentclass[aps,prd,reprint,twocolumn,superscriptaddress,longbibliography,nofootinbib,floatfix,showpacs]{revtex4-1}
\usepackage{epsfig}
\usepackage{amsmath,amssymb,amsfonts}
\usepackage{hyperref}
\usepackage{mathrsfs}
\usepackage{bbm}
\usepackage{slashed}
\usepackage{graphicx}
\usepackage{verbatim}

\usepackage{bm}

\usepackage[usenames]{color}

\usepackage{tikz}
\usetikzlibrary{calc,decorations.markings,decorations.pathmorphing,positioning,shapes,snakes,arrows}

\definecolor{darkgreen}{rgb}{0.2,0.6,0}

\newcommand{\be}{\begin{equation}}
\newcommand{\ee}{\end{equation}}
\newcommand{\bw}{\begin{widetext}}
\newcommand{\ew}{\end{widetext}}
\newcommand{\bi}{\begin{itemize}}
\newcommand{\ei}{\end{itemize}}
\newcommand{\bea}{\begin{eqnarray}}
\newcommand{\eea}{\end{eqnarray}}
\newcommand{\ud}{\mathrm{d}}

\newcommand{\LCm}{{\scriptscriptstyle -}} 
\newcommand{\LCp}{{\scriptscriptstyle +}}

\usepackage[T1]{fontenc} \usepackage[latin1]{inputenc}

\begin{document}
%



\title{Resummation of the $\alpha$ expansion for nonlinear pair production}

\author{Greger Torgrimsson}
\email{greger.torgrimsson@umu.se}
\affiliation{Department of Physics, Ume{\aa} University, SE-901 87 Ume{\aa}, Sweden}

\begin{abstract}

We show how to resum the Furry-picture $\alpha$ expansion in order to take quantum radiation reaction and spin transition into account in the nonlinear trident process in (pulsed) plane-wave background fields. The results are therefore nontrivial functions of both the background field strength, $eE$, and the coupling to the quantized photon field, $\alpha=e^2/4\pi$. The effective expansion parameter, $T$, is $\alpha$ times $eE/m\omega\gg1$, which makes higher orders important. We show that they can change the sign of the spin dependent part already at $T<1$, which will be experimentally accessible. 
	
\end{abstract}	

\maketitle

\section{Introduction}

The nonlinear trident process~\cite{Dinu:2017uoj,King:2018ibi,Mackenroth:2018smh,Acosta:2019bvh,Krajewska15,Hu:2014ooa,King:2013osa,Ilderton:2010wr,Hu:2010ye,Bamber:1999zt,Ritus:1972nf,Baier,Dinu:2019wdw,Torgrimsson:2020wlz} in a strong (laser) background field, $e^\LCm\to e^\LCm e^\LCm e^\LCp$, is an experimentally important process in strong-field QED, see~\cite{DiPiazza:2011tq,Gonoskov:2021hwf,Fedotov:2022ely} for reviews. It was measured in an experiment~\cite{Bamber:1999zt} that was the first and, for a long time, basically the only experiment in this research field. Back then the lasers were actually relatively weak, i.e.\footnote{We absorb $e$ into the laser field strength, $eE\to E$, and use units with $c=\hbar=m_e=1$.} $a_0=E/\omega<1$. Laser intensities have since increased steadily and it is now possible to have $a_0\gg1$. There are plans to measure trident again but now in a genuinely strong-field regime, e.g. by LUXE~\cite{Abramowicz:2021zja} or FACET-II~\cite{Meuren:2020nbw}.  

For $a_0>1$ one cannot treat the background field in perturbation theory. The quantized photon field, though, is still treated in perturbation theory, which gives $P_{\rm trident}=\alpha^2F(E)$ to leading order in $\alpha$, where $F$ is some nontrivial function. This $\mathcal{O}(\alpha^2)$ has been studied in several recent papers~\cite{Dinu:2017uoj,King:2018ibi,Mackenroth:2018smh,Acosta:2019bvh,Krajewska15,Hu:2014ooa,King:2013osa,Ilderton:2010wr,Hu:2010ye,Dinu:2019wdw,Torgrimsson:2020wlz}, and we now have a much better understanding of how to calculate it.

However, for $a_0\gg1$, the effective expansion parameter is $T=a_0\alpha$, which is not small, i.e. $\mathcal{O}(\alpha^2)$ may not be enough. In this paper we will present methods for how to resum all orders in $\alpha$, and show that this is important even for $T\lesssim1$.
Various resummations of the $\alpha$ expansion appear in several recent papers on radiation reaction (RR)~\cite{Heinzl:2021mji,Torgrimsson:2021wcj,Torgrimsson:2021zob,Ekman:2021eqc,Ekman:2021czy} and other processes~\cite{Karbstein:2019wmj,Karbstein:2021gdi,Mironov:2020gbi,Mironov:2022jbg,Edwards:2020npu,Podszus:2021lms,Podszus:2022jia}.

\section{Derivation}

We showed in~\cite{Dinu:2019pau,Torgrimsson:2020gws} how to write the dominant contribution of $\mathcal{O}(\alpha^n)$ processes with products of $\mathcal{O}(\alpha)$ ``strong-field-QED Mueller matrices''. We showed in~\cite{Torgrimsson:2021wcj,Torgrimsson:2021zob} how to evaluate and resum the resulting $\alpha$ expansion for the momentum expectation value or the spin-transition probability of an electron in the laser. Quantum effects become important well before pair production sets in. Here we take the next step and consider the production of one pair. We therefore neglect terms and processes that are more exponentially suppressed than the leading exponential scaling, which is $\exp(-16/[3\chi])$ for a constant field.

Plane waves only depend on lightfront time, $\phi=kx=\omega(t+z)$, so the other coordinates give trivial integrals. And only the lightfront longitudinal\footnote{Since $P_3$ never appears separately in this paper, we drop "lightfront" and simply call $P_\LCm$ the longitudinal momentum.} components of the particle momenta, $P_\LCm=(P_0-P_3)/2>0$, play a nontrivial role, as the perpendicular  integrals factorize and can therefore be performed for each Mueller matrix separately, which has already been done. 

To obtain $\mathcal{O}(\alpha^n)$, we start with the latest vertices and work backwards in time. Since we will not consider the spin or momentum of the final-state particles, the last step is the one where an intermediate photon decays into a pair (see Fig.~\ref{diagramFig}). In this step we only need a Mueller vector rather than matrix, which takes into account the dependence on the polarization of the intermediate photon. Here we use\footnote{Lorentz contractions are denoted simply $kp=k_0p_0-k_jp_j$.} $s_2=kp_2/kl$ and $s_3=kp_3/kl=1-s_2$ for the ratios of the longitudinal momenta of the electron and positron, respectively, and the photon. At each step we let $b_0=kP$, where $P_\mu$ is the momentum of whatever particles that goes into that step, so in this step $b_0=kl$. The Mueller vector we need is given by~\cite{Torgrimsson:2020gws}
\be
{\bf M}^{BW}(\chi,s_3)=\begin{pmatrix}\text{Ai}_1(\xi)-\kappa\frac{\text{Ai}'(\xi)}{\xi}&\frac{\text{Ai}'(\xi)}{\xi} \end{pmatrix} \;,
\ee
where
$\text{Ai}_1(\xi)=\int_\xi^\infty\ud x\,\text{Ai}(x)$,
and\footnote{In the literature, $\xi$ is another common symbol for what we call $a_0$. We never use $\xi$ for $a_0$.} $\xi=(r/\chi)^{2/3}$, $r=(1/s_2)+(1/s_3)$, $\kappa=(s_2/s_3)+(s_3/s_2)$ and $\chi=\chi_0|f'(\phi)|=a_0b_0|f'(\phi)|$ is the locally-constant value of $\chi=\sqrt{-(F^{\mu\nu}l_\nu)^2}$ (the potential is given by $a_\mu(\phi)=\delta_{\mu1}a_0 f(\phi)$). From this we construct a $\sigma$-dependent Stokes vector
\be
{\bf N}^{(1)}(\chi_0,\sigma):=\int_{\sigma}^\infty\frac{\ud\sigma'}{\chi_0}\int_0^1\ud s_3\,{\bf M}^{BW}(\chi,s_3) \;.
\ee
The lower integration limit, $\sigma$, allows us to prepend the earlier Mueller matrices with time-ordering. An initial-state photon with Stokes vector ${\bf N}$ would decay into a pair with probability $P={\bf N}_\gamma\cdot{\bf N}^{(1)}(\chi_0,-\infty)$. In general, Stokes vectors have four elements, but, for the cases we consider here, only two of them are relevant. ${\bf N}_\gamma=\{1,\pm1\}$ and ${\bf N}_\gamma=\{1,0\}$ correspond to photon polarization parallel to the electric and magnetic fields and to an unpolarized photon. 

This photon is produced by an electron. For general spin and polarization, we would need the Mueller matrix in~\cite{Torgrimsson:2020gws} that describes how ${\bf N}_\gamma$ changes before the photon decays. But for the case we consider here this birefringence does not contribute. Thus, the second step is the photon emission step. For this we need a Mueller matrix that takes into account the dependence on the spin of the electron and the polarization of the intermediate photon. This has the structure ${\bf N}_0\cdot{\bf M}^C_{0\gamma}\cdot{\bf N}_\gamma$ where ${\bf N}_0$ is the Stokes vector for the electron before emitting the photon. ${\bf N}_0$ too is reduced from a vector with four to two elements, where ${\bf N}_0=\{1,\pm1\}$ and ${\bf N}_0=\{1,0\}$ correspond to spin (anti-)parallel to the magnetic field and to an unpolarized state. The momentum of the electron before and after emitting the photon is $p_\mu$ and $p_{1\mu}$, respectively, so for this step we let $b_0=kp$. We also use $q=kl/kp$ and $s_1=kp_1/kp=1-q$.
The effectively $2\times2$ Mueller matrix is given by
\be\label{MC0gamma}
{\bf M}^C_{0\gamma}(\chi,q)=\begin{pmatrix}-\text{Ai}_1(\xi)-\kappa\frac{\text{Ai}'(\xi)}{\xi}&-\frac{\text{Ai}'(\xi)}{\xi}\\
q\frac{\text{Ai}(\xi)}{\sqrt{\xi}} &
\frac{q}{s_1}\frac{\text{Ai}(\xi)}{\sqrt{\xi}} \end{pmatrix} \;,
\ee   
where now $\xi=(r/\chi)^{2/3}$, $r=(1/s_1)-1$ and $\kappa=(1/s_1)+s_1$. The second step is given by
\be\label{recursive1to2}
{\bf N}^{(2)}(\chi_0,\sigma)=\int_\sigma^\infty\frac{\ud\sigma'}{\chi_0}\int_0^1\ud q\,{\bf M}^{C}_{0\gamma}(\chi,q)\cdot{\bf N}^{(1)}(q\chi_0,\sigma') \;.
\ee 
Note that we now have $q\chi_0$ instead of $\chi_0$ in the argument of ${\bf N}^{(1)}$ since at each step we use $b_0$ for the momentum of whatever particle that is present just before that step. 

Starting from ${\bf N}^{(3)}$ we have the same recursive formula as in~\cite{Torgrimsson:2021wcj,Torgrimsson:2021zob}, where we studied $\langle kP\rangle$ for an electron that does not produce pairs. For an electron experiencing RR we need two Mueller matrices, one for Compton scatting
\be
{\bf M}^C(\chi,q)=\begin{pmatrix}-\text{Ai}_1(\xi)-\kappa\frac{\text{Ai}'(\xi)}{\xi} & \frac{q}{s_1}\frac{\text{Ai}(\xi)}{\sqrt{\xi}}\\
q\frac{\text{Ai}(\xi)}{\sqrt{\xi}}&-\text{Ai}_1(\xi)-2\frac{\text{Ai}'(\xi)}{\xi} \end{pmatrix}
\ee 
and one for the loop
\be
{\bf M}^L(\chi,q)=\begin{pmatrix}\text{Ai}_1(\xi)+\kappa\frac{\text{Ai}'(\xi)}{\xi} & -q\frac{\text{Ai}(\xi)}{\sqrt{\xi}}\\
-q\frac{\text{Ai}(\xi)}{\sqrt{\xi}}&\text{Ai}_1(\xi)+\kappa\frac{\text{Ai}'(\xi)}{\xi} \end{pmatrix} \;,
\ee
with the same $\xi$ and $\kappa$ as for ${\bf M}^C_{0\gamma}$. From $({\bf M}^C+{\bf M}^L)\cdot\{1,0\}=\{0,0\}$ we can explicitly see that, if we do not consider the momentum or spin of the final-state particles, there is no RR correction on the fermion lines after the emission of the photon that decays into a pair.
For $n\geq3$, we have\footnote{A recursive formula for a different object was obtained in~\cite{Tamburini:2019tzo}.} (cf.~\cite{Torgrimsson:2021wcj,Torgrimsson:2021zob})
\be\label{recursiveGeneral}
\begin{split}
{\bf N}^{(n)}(\chi_0,\sigma)=&\int_\sigma^\infty\frac{\ud\sigma'}{\chi_0}\int_0^1\ud q\{{\bf M}^L(\chi,q)\cdot{\bf N}^{n-1}(\chi_0,\sigma') \\
&+{\bf M}^C(\chi,q)\cdot{\bf N}^{(n-1)}([1-q]\chi_0,\sigma')\} \;.
\end{split}
\ee
The trident probability is obtained by resumming the $\alpha$ expansion,
$P={\bf N}_0\cdot{\bf N}(\chi_0,-\infty)$,
where ${\bf N}_0$ describes the spin of the initial electron and
\be
{\bf N}(\chi_0,\sigma)=\sum_{n=2}^\infty T^n{\bf N}^{(n)}(\chi_0,\sigma) \;,
\ee
where $T=a_0\alpha$. ${\bf N}(\chi_0,\sigma)$ can be obtained either 1) by calculating the first e.g. 10 terms, ${\bf N}^{(2)}$ to ${\bf N}^{(11)}$, and then resumming them with some appropriate method, see below; or 2) by resumming before computing, i.e. solving the following integrodifferential\footnote{Integrodifferential equations, for different objects, also appear in macroscopic, kinetic approaches, see e.g.~\cite{Sokolov:2010am,Elkina:2010up,Neitz:2014hla,Seipt:2020uxv}.} equation 
\be\label{integroDifferential}
\begin{split}
\frac{\partial{\bf N}}{\partial\sigma}=&T^2\frac{\partial{\bf N}^{(2)}}{\partial\sigma}\\
&-T\int_0^1\frac{\ud q}{\chi_0}\{{\bf M}^L\cdot{\bf N}(\chi_0)+{\bf M}^C\cdot{\bf N}([1-q]\chi_0)\} \;.
\end{split}
\ee  
We integrate this backwards in time starting with ${\bf N}(\chi_0,+\infty)=\{0,0\}$.
Note that, while~\eqref{recursiveGeneral} has the same form as Eq.~(1) in~\cite{Torgrimsson:2021wcj}, \eqref{integroDifferential} has an extra, inhomogeneous term compared to Eq.~(2) in~\cite{Torgrimsson:2021wcj}.


For a constant field, the $\sigma$ integrals simply gives $\int\ud\sigma_1...\ud\sigma_n=\Delta\phi^n/n!$, with $n!$ due to time ordering. It is natural to absorb $\Delta\phi$ into $T=\Delta\phi a_0\alpha$. Hence, \eqref{recursiveGeneral} reduces to 
\be\label{recursiveForN}
{\bf N}^{(n)}=\int_0^1\!\frac{\ud q}{n\chi}\{{\bf M}^C\cdot{\bf N}^{(n-1)}(\chi[1-q])+{\bf M}^L\cdot{\bf N}^{(n-1)}(\chi)\} \;.
\ee
We can now use $T$ rather than $\sigma$ as variable for an integrodifferential equation,
\be\label{integroDiffT}
\begin{split}
&\frac{\partial}{\partial T}{\bf N}(T,\chi)=2T{\bf N}^{(2)}(\chi)\\
&+\int_0^1\frac{\ud q}{\chi}\left\{{\bf M}^C\cdot{\bf N}(T,\chi[1-q])+{\bf M}^L\cdot{\bf N}(T,\chi)\right\} \;,
\end{split}
\ee
with ``initial'' condition ${\bf N}(0,\chi)=\{0,0\}$. 

\begin{figure}
\includegraphics[width=\linewidth]{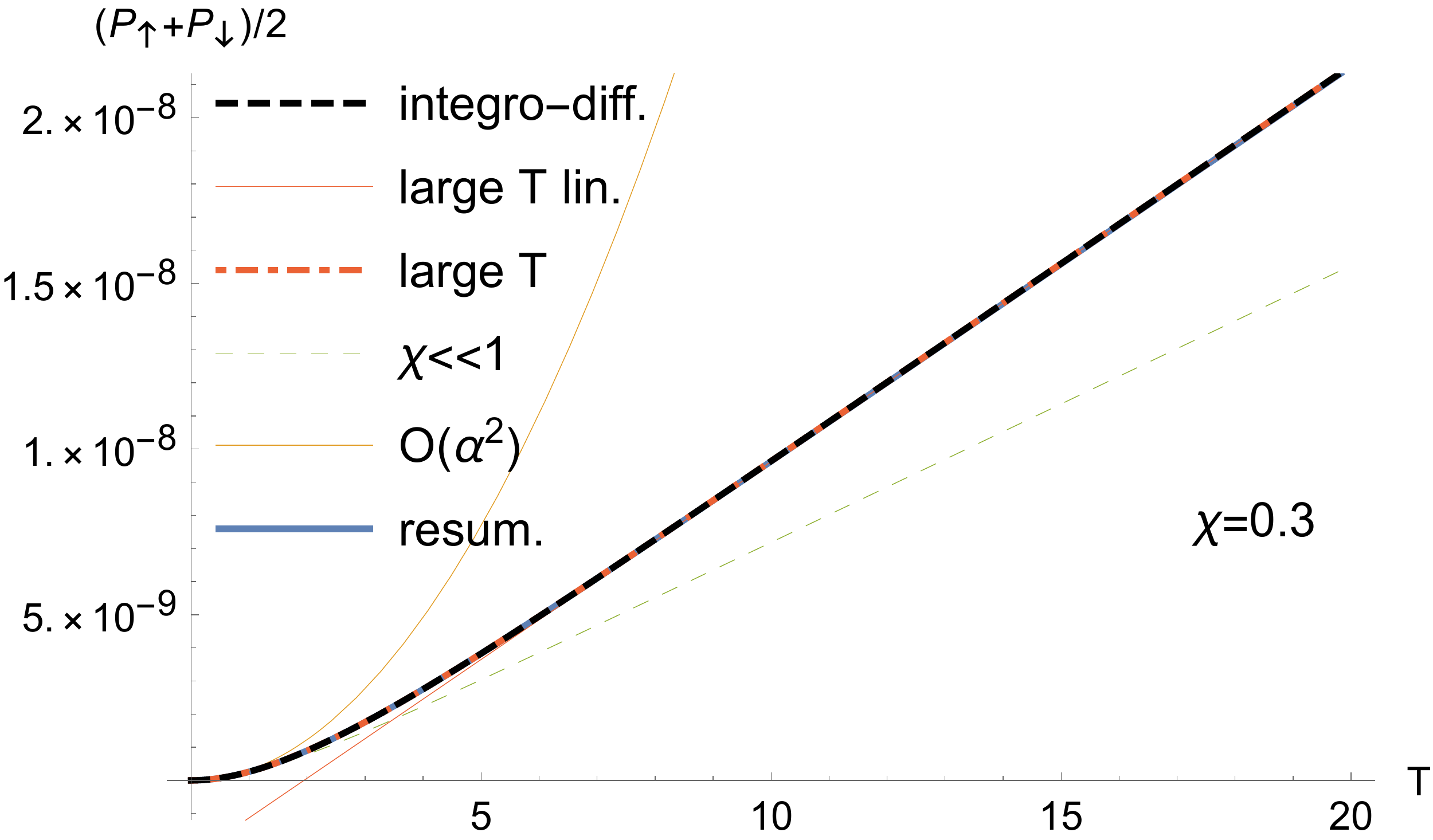}
\caption{The initial electron has spin up or down along the magnetic field, $P_\downarrow$ and $P_\uparrow$. The ``resum.'' line is obtained by resumming the $\chi$ expansion with Pad\'e-Borel and the $\alpha$ expansion with the resummation method in~\eqref{Hresummation}, with $n=1$ in~\eqref{PadeHgen}. The ``integro-diff.'' line is a solution to~\eqref{integroDiffT}. The $\mathcal{O}(\alpha^2)$ line gives the trident probability with no RR. The $\chi\ll1$ line gives the low energy limit in~\eqref{PaveClassical}. The ``large T'' lines are obtained in Appendix~\ref{large T limit}. See Fig.~\ref{chi03plotDiff} for $P_\downarrow-P_\uparrow$.}
\label{chi03plots}
\end{figure}

\begin{figure}
\includegraphics[width=\linewidth]{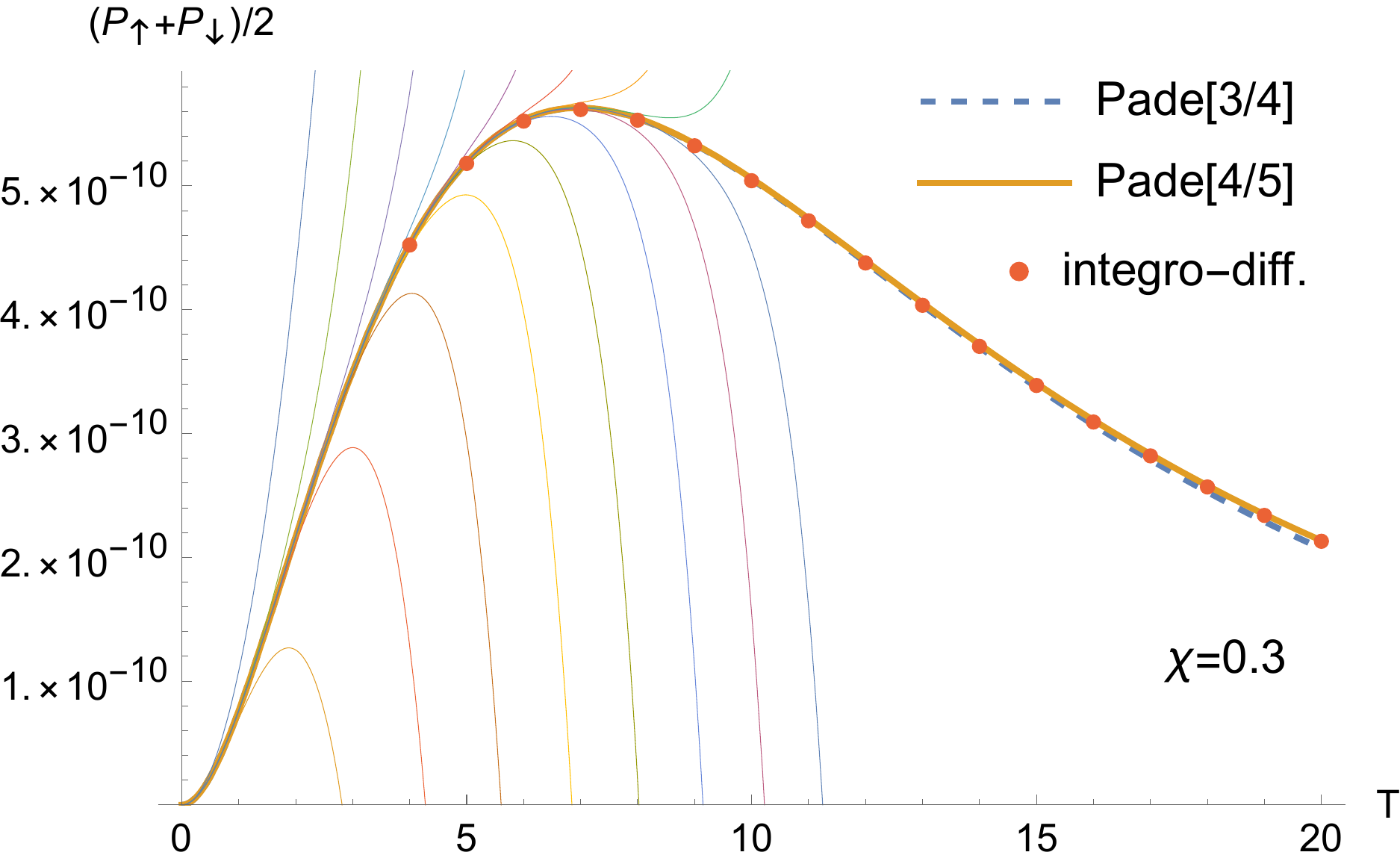}
\includegraphics[width=\linewidth]{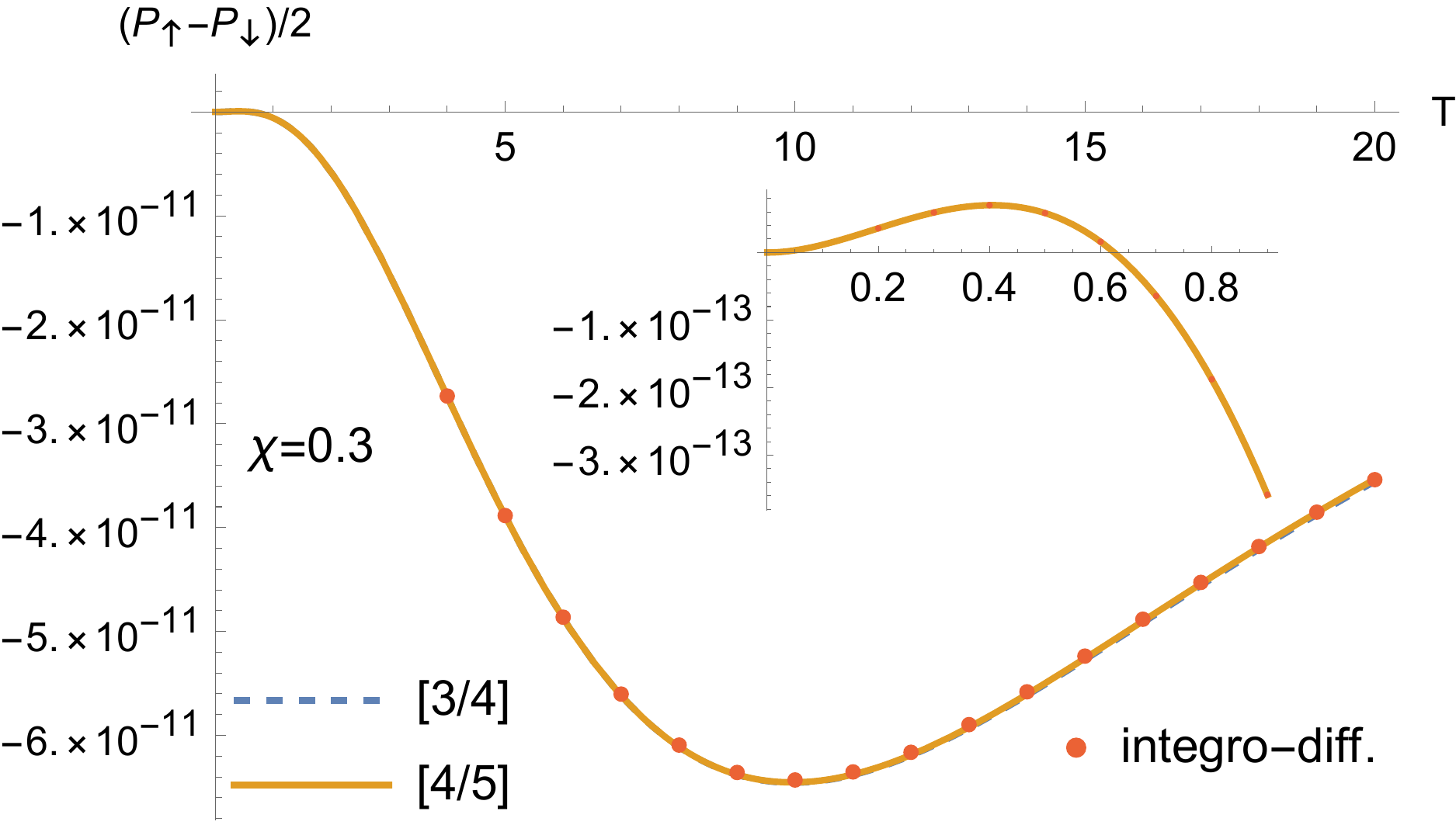}
\caption{Similar to Fig.~\ref{chi03plots} but for a Sauter pulse. The thin lines for $(P_\downarrow+P_\uparrow)/2$ show the results of just adding the $n$ terms $\mathcal{O}(\alpha^2)$ to $\mathcal{O}(\alpha^{n+1})$ without any resummation. The $[m/m+1]$ lines are obtained with the method in Sec.~\ref{Resumming convergent series}, and the dots are solutions to~\eqref{integroDifferential}.}
\label{SauterPlots}
\end{figure}

The final results are shown in Fig.~\ref{chi03plots}, \ref{SauterPlots} and~\ref{chi5plots}, which shows perfect agreement between the results obtained from~\eqref{integroDifferential} or~\eqref{integroDiffT} and from~\eqref{Hresummation}. 

\section{Leading contribution in $\chi\ll1$}\label{low energy constant}

Here we obtain the leading order in $\chi\ll1$ for a constant field. We start with the ansatz
\be
{\bf N}^{(n)}=\{a_n+\chi c_n,\chi b_n\}\exp\left(-\frac{16}{3\chi}\right) \;,
\ee
where $a_n$ and $b_n$ are constants. At $n=2$, we have 
$a_2=\frac{1}{2}\frac{1}{32}$ and $b_2=\frac{1}{2}\frac{1}{32\times27}$.
It turns out that we do not need $c_n$ in order to obtain $a_n$ and $b_n$. 
From~\eqref{recursiveForN} 
we find (see appendix)
\be\label{anFromnm1}
a_n=-\frac{d}{n}a_{n-1} 
\qquad
b_n=-\frac{1}{n}[f a_{n-1}+d b_{n-1}] \;,
\ee
where $d=\mathcal{J}_{\text{Ai}_1}(0)+2\mathcal{J}_{\text{Ai}'}(0)-\mathcal{I}_{\text{Ai}_1}(0)-2\mathcal{I}_{\text{Ai}'}(0)\approx0.711201$ and $f=\mathcal{I}_\text{Ai}(1)-\mathcal{J}_\text{Ai}(1)\approx0.419148$, and so
\be\label{anFactorial}
\begin{split}
a_n&=\frac{2(-d)^{n-2}}{n!}a_2 \\
b_n&=\frac{2(-d)^{n-2}}{n!}\left(b_2+\frac{f}{d}(n-2)a_2\right) \;.
\end{split}
\ee
Thus, averaging over the initial spin,
\be\label{PaveClassical}
\begin{split}
\langle P\rangle&=\{1,0\}\cdot\sum_{n=2}^\infty T^n{\bf N}^{(n)}\approx
e^{-\frac{16}{3\chi}}\sum_{n=2}^\infty a_n T^n \\
&=\frac{T^2F(dT)}{2\times32}\exp\left(-\frac{16}{3\chi}\right) \;,
\end{split}
\ee
where $F(x)=\frac{2}{x^2}\left[e^{-x}-1+x\right]$.
For the spin difference we find
\be\label{PdiffClassical}
\begin{split}
\frac{P_{\uparrow}-P_{\downarrow}}{2}&=\{0,1\}\cdot\sum_{n=2}^\infty T^n{\bf N}^{(n)}\approx
\chi e^{-\frac{16}{3\chi}}\sum_{n=2}^\infty b_nT^n\\
&=\left(\frac{1}{27}F(dT)-\frac{f}{d}G(dT)\right)\frac{T^2}{64}\chi\exp\left(-\frac{16}{3\chi}\right) \;,
\end{split}
\ee
where $G(x)=\frac{2}{x^2}[x-2+(2+x)e^{-x}]$.
If $a_0\Delta\phi$ is not large, i.e. $T\ll1$, we have $F(0)=1$ and the resummation reduces to trident at leading $\mathcal{O}(\alpha^2)$. However, if $a_0\Delta\phi$ is large enough to not only compensate for the factor of $\alpha$ in $T$ but to make $T\gg1$, then we have $T^2F(dT)\to2T/d$, i.e. the probability grows linearly, $P\sim T$. We can understand this as follows. Once the intermediate photon has been emitted, it can decay anywhere in the field, which gives a temporal volume factor $\sim T$. Without RR, the electron can emit the intermediate photon anywhere in the pulse, which gives another factor of $T$. However, with RR, the electron's longitudinal momentum decreases over time, so the electron can only emit (with significant probability) a sufficiently high-energy photon during a limited time interval, i.e. there is no additional factor of $T$.


We find that $P_{\uparrow}-P_{\downarrow}$ changes sign as $T$ increases, and that this happens already at $T\sim0.3$. Thus, from the $\mathcal{O}(\alpha^2)$ results we have that $P_{\uparrow}>P_{\downarrow}$ for $T\ll1$, but as $T$ increases we instead find $P_{\downarrow}>P_{\uparrow}$, and $a_0$ actually does not have to be extremely large for this to happen.

\section{Resumming convergent series}\label{Resumming convergent series}

In~\cite{Torgrimsson:2021wcj,Torgrimsson:2021zob} we found $\alpha$ expansions with finite radius of convergence, which we therefore resummed with Pad\'e approximants.
In contrast, in Sec.~\ref{low energy constant} we see an infinite radius.
In principle one can sum such series directly without any resummation. But that would mean having to calculate more and more terms to reach convergence as we increase $T$ (see Fig.~\ref{SauterPlots}). This is neither efficient nor practical, because, in contrast to Sec.~\ref{low energy constant}, we will in general only be able to obtain a finite number of terms, say up to $\mathcal{O}(\alpha^{n_{\rm max}})$.  A direct sum,
$\sum_{n=0}^{n_{\rm max}}T^n{\bf N}^{(n)}$, scales as $T^{n_{\rm max}}$ as $T\to\infty$, which is not physical since $n_{\rm max}$ is just the order where we happened to stop. Thus, we still need to resum this type of series.

Recall first that an asymptotic series can be resummed with the Borel-Pad\'e approach where one first divides the coefficients by $n!$ and then forms a Pad\'e approximant (see Appendix~\ref{Borel resummation}). As the coefficients go as $|c_n|\sim1/n!$ we will instead multiply by $n!$. So, we insert a factor of
\be
1=n!\int_\gamma\frac{\ud t}{2\pi i}e^t t^{-(1+n)} \;,
\ee
where $\gamma$ starts at $t=-\infty-i\epsilon$, wraps around the negative real axis and ends at $t=-\infty+i\epsilon$. See Eq.~(5.9.2) in~\cite{HankelDLMF} for this ``Hankel's loop integral''. We can then write\footnote{We refrain from calling this the ``Hankel transform'' since that name is already used for something else.}
\be
\psi(x)=\int_\gamma\frac{\ud t}{2\pi i}\frac{e^t}{t}H\psi(x/t) 
\qquad
H\psi(z)=\sum_{n=0}^\infty n!c_n z^n \;.
\ee
Since $|c_n|\sim1/n!$ at large $n$, $H\psi(z)$ has a finite radius of convergence. We can therefore resum the truncated transform
by matching it onto a Pad\'e approximant. The final resummation is thus given by
\be\label{Hresummation}
\psi(x)=\int_\gamma\frac{\ud t}{2\pi i}\frac{e^t}{t}PH\psi(x/t) \;,
\ee
where the integral can be performed with the residue theorem. The final result is thus a sum of products of polynomials and exponentials ($e^{-\text{const.}T}$) similar to the low-energy limit in~\eqref{PaveClassical} and~\eqref{PdiffClassical}. 
If one can guess some appropriate order of these polynomials and the number of different exponentials, then one can of course obtain the coefficients by directly matching with the $T$ expansion, i.e. without introducing the Hankel integral. However, for the examples we have tried, it seems to be much easier to obtain a good resummation by first making this transformation. 

Recall that Pad\'e approximants can have spurious poles, and note that, in general, we can only obtain the $\alpha$ expansion to a finite precision. We have in some cases found that this method can give terms with $e^{+\text{const.}T}$, where $\text{Re}(\text{const.})>0$, but with a numerically very small pre-exponential factor which makes this unphysical $T$ scaling nevertheless negligible for reasonably large $T$. 
For larger $T$ one can try to fix such cases by simply removing the terms with $e^{+\text{const.}T}$. If the corresponding pre-exponential coefficients are several orders of magnitude smaller than the coefficients in front of terms with $e^{-\text{const.}T}$, then one can expect that the $T\ll1$ expansion of $\psi_{\rm fix}$ is still correct to a good precision\footnote{Without this fix, the expansion of $\psi_{\rm resum}$ will agree with ${\bf N}^{(n)}$, $n\leq n_{\rm max}$, to within the working precision (e.g. $10^{-15}$). But if we only know ${\bf N}^{(n)}$ with a precision of e.g. $10^{-5}$, then it is not a problem if $\psi_{\rm fix}$ only agrees with ${\bf N}^{(n)}$ to a precision of $10^{-5}$. While this fix seems to work well, it still leaves some inspiration for trying to find more optimal use of the $n_{\rm max}$ terms calculated.}. In any case, this has not been a problem for the cases shown in the plots.

To check that this works, consider first
$\psi=\sum_{n=2}^\infty a_nT^n$
with $a_n$ given by~\eqref{anFactorial}. For this example we find a simple geometric series for $H\psi$,
\be\label{geometricHpsi}
H\psi=2a_2T^2\sum_{n=2}^\infty\frac{(-dT)^{n-2}}{t^n}=\frac{2a_2T^2}{t(t+dT)} \;.
\ee
The radius of convergence of this series is $|dT/t|<1$, so it makes sense to choose the integration contour $\gamma$ such that $|t|>dT$. We can now perform the $t$ integral with the residue theorem. We have poles at $t=0$ and $t=-dT$, and they both contribute. We find
\be
\psi(x)=\int_\gamma\frac{\ud t}{2\pi i}\frac{e^t}{t}H\psi=a_2T^2 F(dT) \;,
\ee
which agrees with~\eqref{PaveClassical}. We also recover~\eqref{PdiffClassical} in the same way.
In these examples we have access to all terms and we find geometric series that can be resummed as in~\eqref{geometricHpsi}, which is already exactly the ratio of two polynomials. The point is that in general we will not find a geometric series, but we can still resum the $H\psi$ series with Pad\'e approximants. 

For a constant field, the leading order at $T\ll1$ (trident without RR) scales as $T^2$ and we expect the resummed result to scale linearly in $T$ for large $T$, so we choose $[n+1/n]$ Pad\'e approximants as
\be\label{PadeHgen}
PH\psi(z)=\frac{\sum_{i=2}^{n+1}A_iz^i}{1+\sum_{j=1}^n B_jz^j} \;.
\ee
Results are shown in Fig.~\ref{chi03plots}. The convergence is very fast. For $\chi=0.3$ we only need $n=1$ and $n=2$ for the unpolarized and polarized parts, i.e. we only need terms up to $\mathcal{O}(\alpha^3)$ and $\mathcal{O}(\alpha^5)$. Note that for $n=1$ we have exactly the functional dependence of $T$ as in~\eqref{PaveClassical}, only the overall coefficient and the constants $d$ are different. There is though no a priori reason to expect that $n=1$ and $n=2$ would be enough if we consider larger $\chi$. But it turns out that we actually have to increase $\chi$ significantly to see this, see appendix. 


\section{Results for a Sauter pulse}

As an example of a pulsed field, we consider a Sauter pulse\footnote{Oscillations tend to average out spin effects. This can be avoided by making the oscillations asymmetric~\cite{Seipt:2019ddd,Chen:2019vly}, or by using dense electron beams to generate non-oscillating fields~\cite{Song:2021wou}.} $a_\mu(\phi)=\delta_{\mu1}a_0\tanh(\phi)$. The results are shown in Fig.~\ref{SauterPlots}.
We have used~\eqref{recursiveGeneral} to obtain the first $\gtrsim10$ terms and resummed them using the method described in Sec.~\ref{Resumming convergent series}.
As for the constant field, for the Sauter pulse, too, we find that $(P_\uparrow-P_\downarrow)/2$ changes sign as $T$ increases and that this happens already for $T<1$.  

In contrast to the constant-field case, here $P$ decreases as $T\to\infty$. It is therefore better to choose $[n/n+1]$ approximants rather than~\eqref{PadeHgen}, to remove the pole at $t=0$, which would otherwise give terms without exponential suppression. The results agree with the solution to~\eqref{integroDifferential}.
We can understand the different asymptotic scaling roughly as follows. We model the exponential suppression of pair production by $e^{-\text{const.}/\chi(\phi)}$, where $\chi(\phi)=kP(\phi)a_0f'(\phi)$ is the product of a local field strength $a_0f'(\phi)$ and a local momentum $kP(\phi)$, which we  estimate using the solution to the Landau-Lifshitz equation~\cite{DiPiazzaLLsol,HeintzmannGrewingLLsol}, $kP(\phi)\to b_0/(1+[2/3]Ta_0\int_{-\infty}^\phi\ud\sigma\,f'(x))$. $kP(\phi)$ decreases and hence favors production early in the pulse, while $a_0f'(\phi)$ favors production close to the field maximum. For a constant field, $f'=1$, only $kP(\phi)$ is relevant and the dominant contribution comes from the time just after the electron has entered the field and before it has lost too much momentum, which gives a $T$-independent exponent, $e^{-\text{const.}/\chi(-\infty)}$. For a Sauter pulse, $f'(\phi)=\text{sech}^2(\phi)$, the dominant contribution comes from $\chi'(\phi_d)=0$, giving $\phi_d=-(1/4)\ln[1+(4/3)a_0b_0T]$, and $e^{-\text{const.}/\chi(\phi_d)}\sim e^{-\text{const.}'T}$ as $T\to\infty$.

\section{Conclusions and outlook}

We have derived recursive and integrodifferential matrix equations that give quantum RR to all orders in $\alpha$ for the probability of nonlinear trident. We have shown that corrections to the leading order (no RR) become important already for values of $T=a_0\alpha$ that can be generated with today's lasers. We have also shown how to resum convergent $\alpha$ expansions.

Here we have focused on $a_0\gg1$ for which the Mueller matrices can be expressed in terms of Airy functions. This method, though, can be used even if $a_0\sim1$, provided one uses the appropriate Mueller matrices in~\cite{Dinu:2019pau,Torgrimsson:2020gws} and the pulse is long. For general spin/polarization, one may have to include a resummation of the Mueller matrix in~\cite{Torgrimsson:2020gws} that takes birefringence of the intermediate photon into account.
It would also be interesting to consider e.g. the spin of the produced positron. This has been studied using PIC codes in~\cite{Chen:2019vly,Li:2020bwo,Song:2021wou}. For that we would need to take RR into account on the positron line, i.e. with ${\bf M}^C$ and ${\bf M}^L$. When planning such generalizations, it is encouraging to note that we have been able to resum the $\alpha$ expansions both in this paper and in~\cite{Torgrimsson:2021wcj,Torgrimsson:2021zob} with relatively few terms.

\acknowledgements

G.~T. is supported by the Swedish Research Council, contract 2020-04327.

\appendix

\section{$\chi$-expansion approach}

In this section we will explain how to obtain the constant-field results by making a second expansion, i.e. by expanding each order in $\alpha$ in an asymptotic expansion in $\chi$.
We work backwards, starting with the pair-production step.

To obtain the $\chi$ expansions we need the following expansions of the Airy functions. Let $\gamma=r/\chi$. For large $\gamma$ we can obtain an expansion of $\text{Ai}_1(\gamma^{2/3})$ by first writing it in terms of the following integral representation
\be
\text{Ai}_1(\gamma^{2/3})=\frac{i}{2\pi}\int_\infty^\infty\frac{\ud\tau}{\tau}\exp\left\{i\gamma\left(\tau+\frac{\tau^3}{3}\right)\right\} \;,
\ee
where the integration contour passes above the pole. We can now obtain an expansion using the saddle-point method, i.e. we change variable from $\tau=i+(1/\sqrt{\gamma})\delta\tau$ to $\delta\tau$, expand the integrand in a series in $1/\gamma$ and perform the resulting Gaussian integrals. We find
\be
\text{Ai}_1(\gamma^{2/3})=\frac{\exp\left(-\frac{2\gamma}{3}\right)}{2\sqrt{\pi\gamma}}
\left(
1-\frac{41}{48 \gamma }+\frac{9241}{4608
	\gamma ^2}+\dots
\right) \;,
\ee
where we can quickly obtain the next $>100$ terms. The corresponding expansion for $\text{Ai}'(\gamma^{2/3})/\gamma^{2/3}$ can be obtained directly from the known expansion of the Airy function, one finds
\be
\frac{\text{Ai}'(\gamma^{2/3})}{\gamma^{2/3}}=-\frac{\exp\left(-\frac{2\gamma}{3}\right)}{2\sqrt{\pi\gamma}}
\left(
1+\frac{7}{48 \gamma }-\frac{455}{4608
	\gamma
	^2}+\dots
\right) \;.
\ee
To obtain an expansion for the pair-production probability integrated over the longitudinal momentum, we first change variables from $s_3$ to $r=(1/[1-s_3])+(1/s_3)$. Due to $\exp(-2r/[3\chi])$ the integrand can be expanded around the minimum of $r$, which is $r=4$. We therefore change variable from $r=4+\chi R^2$ to $R$. Expanding the integrand in a series in $\chi$ gives integrals on the form
\be
\int_0^\infty\ud R\, R^n\exp\left(-\frac{2}{3}R^2\right)=\frac{\Gamma\left[\frac{3}{2}+\frac{n}{2}\right]}{(2/3)^{(1+n)/2}(1+n)} \;.
\ee
We thus obtain
\be\label{NBW}
\begin{split}
	T{\bf N}^{(1)}=&
	T\frac{3}{16} \sqrt{\frac{3}{2}}\exp\left(-\frac{8}{3\chi}\right) \\
	\times&\left\{
	1-\frac{11 \chi }{64}+\frac{7985 \chi
		^2}{73728}+\dots\right., \\
	&\left.   
	-\frac{1}{3}+\frac{65 \chi
	}{576}-\frac{21361 \chi ^2}{221184}\dots\right\}
\end{split} \;.
\ee
The probability for nonlinear Breit-Wheeler pair production is given by $P={\bf N}_\gamma\cdot T{\bf N}^{(1)}$, where $T=\alpha a_0\Delta\phi$.
To leading order we recognize the fact that a perpendicularly polarized photon gives twice as large probability compared to a parallel photon, i.e. $\{1,-1\}\cdot{\bf N}^{(1)}\approx2\{1,1\}\cdot{\bf N}^{(1)}$, see~\cite{Reiss62,Nikishov:1964zza,Ritus1985} for the constant-crossed field and~\cite{Esposti:2021wsh} for a general pulsed plane wave.

Now that we have obtained ${\bf N}^{(1)}$, the next step is to prepend ${\bf M}_{0\gamma}^C$ (cf.~\eqref{recursive1to2}) and calculate a corresponding expansion. $\chi=a_0kl$ in~\eqref{NBW} where $l_\mu$ is the intermediate photon momentum. When we prepend ${\bf M}_{0\gamma}^C$ we change notation by replacing $\chi=a_0kl=a_0(kl/kp)kp\to q\chi$, where now $\chi=a_0kp$ and $p_\mu$ is the momentum of the electron before emitting the intermediate photon. The intermediate photon needs to have sufficiently high energy in order to produce a pair, so the probability to emit such a photon also has an exponential expansion similar to~\eqref{NBW}. We obtain this using~\eqref{NBW} and an expansion of~\eqref{MC0gamma}. For the exponential part of the $q$ integral we have
\be
\exp\left\{-\frac{2}{3}(\gamma_2+\gamma_1)\right\} \;,
\ee
where $\gamma_1=4/(q\chi)$ comes from~\eqref{NBW} and from~\eqref{MC0gamma} we have $\gamma_2=r/\chi$ with $r=(1/s_1)-1$ and $s_1=1-q$. There is a saddle point at $q=2/3$, which corresponds to the point where all three final-state fermions have the same momentum, i.e. $s_1=s_2=s_3=1/3$ and $q=1-s_1=s_2+s_3$. In principle we could change variable from $q=(2/3)+\sqrt{\chi}\delta q$ and expand the integrand in a series in $\chi$. However, to obtain a large number of terms in the $\chi$ expansion, it seems faster to instead change variables from
\be
q=\frac{24+3\chi W^2+\sqrt{3\chi}W\sqrt{16+3\chi W^2}}{6(6+\chi W^2)}
\ee
to $W$, where $W(q=0)=-\infty$ and $W(q=1)=+\infty$, which is useful because then the exponent becomes exactly Gaussian
\be
\exp\left\{-\frac{2}{3}(\gamma_2+\gamma_1)\right\}=\exp\left\{-\frac{16}{3\chi}-W^2\right\} \;,
\ee
which means we do not have to expand the exponential part of the integrand in a series in $\chi$. We thus obtain
\be\label{N2chiExpansion}
\begin{split}
	T^2{\bf N}^{(2)}=&\frac{T^2}{2}\frac{\exp\left(-\frac{16}{3\chi}\right)}{32}\\
	\times&\left\{1+\frac{31 \chi }{216}-\frac{3871
		\chi ^2}{31104}+\dots,\right.\\
	&\left.\frac{\chi
	}{27}-\frac{37 \chi ^2}{972}+\dots\right\} \;.
\end{split}
\ee
The probability of trident pair production to leading order in $\alpha$ is given by $P={\bf N}_0\cdot T^2{\bf N}^{(2)}$, where ${\bf N}_0$ is the Stokes vector of the initial electron. The expansion of the unpolarized part, i.e. $\{1,0\}\cdot{\bf N}^{(2)}$, agrees with what we found in~\cite{Torgrimsson:2020wlz}. In order to go beyond the leading order in $\alpha$ we also need the part that describes the dependence on the spin of the initial electron, i.e. $\{0,1\}\cdot{\bf N}^{(2)}$. The leading term in this part, i.e. the one proportional to $\chi/27$, agrees with Eq.~(24) in~\cite{Ritus:1972nf} and Eq.~(92) in~\cite{Dinu:2019pau}\footnote{Our $\hat{B}$ corresponds to $-{\bf e}_2$~\cite{Dinu:2019pau} as explained in Eq.~(52) in~\cite{Torgrimsson:2020gws}.}. Here we have calculated the first $\sim100$ terms in the $\chi$ expansion. 

We obtain the $\chi$ expansions of $\mathcal{O}(\alpha^3)$ and higher orders in $\alpha$ using~\eqref{recursiveForN} with the $\chi$ expansion of ${\bf N}^{(2)}$ as input.
To obtain the $\chi\ll1$ expansion of these orders we need the following integrals. 
We change variables in~\eqref{recursiveForN} from $q=\chi\gamma/(1+\chi\gamma)$ to $\gamma=r/\chi$, where $r=(1/s_1)-1$ and $s_1=1-q$.
From terms with ${\bf M}^L$ we have the same integrals as in~\cite{Torgrimsson:2021wcj}, i.e.
\be
\begin{split}
\mathcal{I}_\text{Ai}(n)&=
\int_0^\infty\ud\gamma\,\gamma^n\frac{\text{Ai}(\gamma^{2/3})}{\gamma^{1/3}}\\
&=\frac{3^{\frac{1}{2}+n}}{4\pi}\Gamma\left[\frac{1}{3}+\frac{n}{2}\right]\Gamma\left[\frac{2}{3}+\frac{n}{2}\right] \;,
\end{split}
\ee
\be
\begin{split}
\mathcal{I}_{\text{Ai}'}(n)&=
\int_0^\infty\ud\gamma\,\gamma^n\frac{\text{Ai}'(\gamma^{2/3})}{\gamma^{2/3}}\\
&=-\frac{3^{\frac{1}{2}+n}}{4\pi}\Gamma\left[\frac{1}{6}+\frac{n}{2}\right]\Gamma\left[\frac{5}{6}+\frac{n}{2}\right] \;,
\end{split}
\ee
\be
\begin{split}
\mathcal{I}_{\text{Ai}_1}(n)&=
\int_0^\infty\ud\gamma\,\gamma^n\text{Ai}_1(\gamma^{2/3})\\
&=\frac{3^{\frac{1}{2}+n}}{2\pi(1+n)}\Gamma\left[\frac{5}{6}+\frac{n}{2}\right]\Gamma\left[\frac{7}{6}+\frac{n}{2}\right] \;.
\end{split}
\ee
From terms with ${\bf M}^C$ we have $\exp(-16/[3(1-q)\chi])=\exp(-16/[3\chi]-16\gamma/3)$, which leads to the following integrals
\be
\begin{split}
&\left\{\mathcal{J}_{\text{Ai}_1},
\mathcal{J}_\text{Ai},\mathcal{J}_{\text{Ai}'}\right\}\\
&=\int_0^\infty\ud\gamma\,\gamma^n e^{-c\gamma}\left\{\text{Ai}_1(\gamma^{2/3}),\frac{\text{Ai}(\gamma^{2/3})}{\gamma^{1/3}},\frac{\text{Ai}'(\gamma^{2/3})}{\gamma^{2/3}}\right\} \;,
\end{split}
\ee
where $c=16/3$. With
\be
\begin{split}
&\left\{\text{Ai}_1(\gamma^{2/3}),\frac{\text{Ai}(\gamma^{2/3})}{\gamma^{1/3}},\frac{\text{Ai}'(\gamma^{2/3})}{\gamma^{2/3}}\right\}\\
&=\int\frac{\ud\tau}{2\pi}\left\{\frac{i}{\tau},1,i\tau\right\}\exp\left[i\gamma\left(\tau+\frac{\tau^3}{3}\right)\right]
\end{split}
\ee
we find
\be
\begin{split}
&\left\{\mathcal{J}_{\text{Ai}_1},
\mathcal{J}_\text{Ai},\mathcal{J}_{\text{Ai}'}\right\}\\
&=n!\int\frac{\ud\tau}{2\pi}\left\{\frac{i}{\tau},1,i\tau\right\}\left[c-i\left(\tau+\frac{\tau^3}{3}\right)\right]^{-(1+n)} \;.
\end{split}
\ee
These integrals can now be performed with the residue theorem. We close the contour in the upper-half complex plane, where there is one pole at
\be
\tau_p=i\left[(8+3\sqrt{7})^{1/3}+\frac{1}{(8+3\sqrt{7})^{1/3}}\right] \;.
\ee
To simplify the calculation of the residue for large $n$, we first perform partial integration
\be
\begin{split}
&\left\{\mathcal{J}_{\text{Ai}_1},
\mathcal{J}_\text{Ai},\mathcal{J}_{\text{Ai}'}\right\}\\
&=\int\frac{\ud\tau}{2\pi}\left[c-i\left(\tau+\frac{\tau^3}{3}\right)\right]^{-1}
\left[\frac{\partial}{\partial\tau}\frac{i}{(1+\tau^2)}\right]^n\left\{\frac{i}{\tau},1,i\tau\right\}\\
&=\mathcal{J}_\text{Ai}(n=0)\left[\frac{\partial}{\partial\tau}\frac{i}{(1+\tau^2)}\right]^n\left\{\frac{i}{\tau},1,i\tau\right\}\bigg|_{\tau=\tau_p} \;,
\end{split}
\ee
where $[...]^n$ means $[...][...]\dots[...]$ with the derivatives acting on everything on the right. For $n=0,1,2,\dots$ we have
\be\label{JintDecimal}
\begin{split}
\mathcal{J}_{\text{Ai}_1}&=
\{0.0458131,0.00685688,0.00211075,\dots\} \\
\mathcal{J}_\text{Ai}&=
\{0.133495,0.0138645,0.00368465,\dots\} \\
-\mathcal{J}_{\text{Ai}'}&=
\{0.388994,0.0225791,0.00518418,\dots\}
\;.
\end{split}
\ee
These numbers can actually be expressed as the roots of third-order polynomials with integer coefficients, e.g.
\be
-1+3\mathcal{J}_\text{Ai}(0)+252\mathcal{J}_\text{Ai}^3(0)=0 \;,
\ee
but it is faster to express them in decimal form. Since precision is often lost in the resummations we are doing, we start with many more digits than those presented in~\eqref{JintDecimal}. 

$\left\{\mathcal{J}_{\text{Ai}_1},
\mathcal{J}_\text{Ai},\mathcal{J}_{\text{Ai}'}\right\}$ grow factorially fast as $n\to\infty$. To obtain this limit we write
\be
\begin{split}
&\left\{\mathcal{J}_{\text{Ai}_1},
\mathcal{J}_\text{Ai},\mathcal{J}_{\text{Ai}'}\right\}
=n!\int\frac{\ud\tau}{2\pi}\left\{\frac{i}{\tau},1,i\tau\right\}\\
&\times\exp\left\{-(1+n)\ln\left[c-i\left(\tau+\frac{\tau^3}{3}\right)\right]\right\} 
\end{split}
\ee
and then perform the integral with the saddle-point method, i.e. we change variable from $\tau=i+(1/\sqrt{n})\delta\tau$ to $\delta\tau$ and expand the integrand in a series in $1/n$. We obtain
\be
\begin{split}
&\hspace{-2.5cm}\left\{\mathcal{J}_{\text{Ai}_1},
\mathcal{J}_\text{Ai},-\mathcal{J}_{\text{Ai}'}\right\}\\
=
\frac{\Gamma
   \left(n+\frac{1}{2}\right)}{2
   \sqrt{6 \pi }6^n}
\bigg\{&
1-\frac{41}{8 n}+\frac{8913}{128
   n^2}-\frac{4635593}{3072 n^3}+\dots,\\
&1-\frac{5}{8 n}+\frac{345}{128
   n^2}-\frac{67085}{3072 n^3}
   +\dots,\\
&1+\frac{7}{8 n}-\frac{399}{128
   n^2}+\frac{73927}{3072 n^3}+\dots   
\bigg\} \;.
\end{split}
\ee

Thus, starting with~\eqref{N2chiExpansion} and repeatedly using~\eqref{recursiveForN} we find, for $n=3$ up to some $n_{\rm max}$ where we decide to stop,
\be
{\bf N}^{(n)}=\exp\left(-\frac{16}{3\chi}\right)\left\{\sum_{m=0}^{m_{\rm max}}a_m\chi^m,\sum_{m=1}^{m_{\rm max}}b_m\chi^m\right\} \;,
\ee
where the coefficients grow factorially with alternating sign, $a_m,b_m\propto(-1)^m m!$ for $m\to\infty$. Note that the $\chi$ expansion of ${\bf N}^{(n)}$ is obtained by inserting the un-resummed $\chi$ expansion of ${\bf N}^{(n-1)}$ into~\eqref{recursiveForN}. 
We have calculated $n_{\rm max}=\mathcal{O}(10)$ and $m_{\rm max}=\mathcal{O}(100)$ terms.
We first resum the $\chi$ expansion of each order in $\alpha$, before we resum the $\alpha$ expansion. 

\section{Borel resummation}\label{Borel resummation}

\begin{figure}
\includegraphics[width=\linewidth]{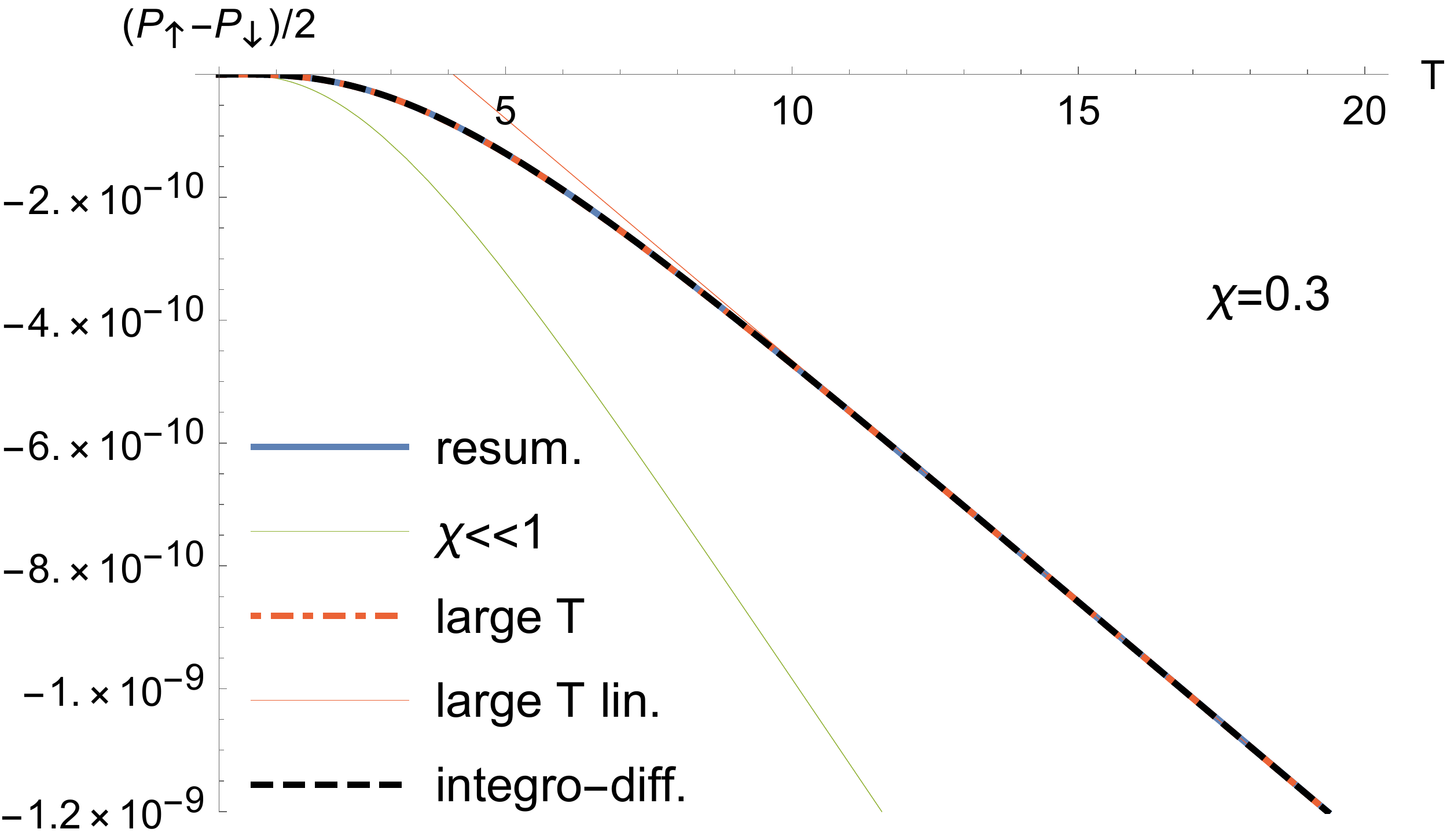}
\includegraphics[width=\linewidth]{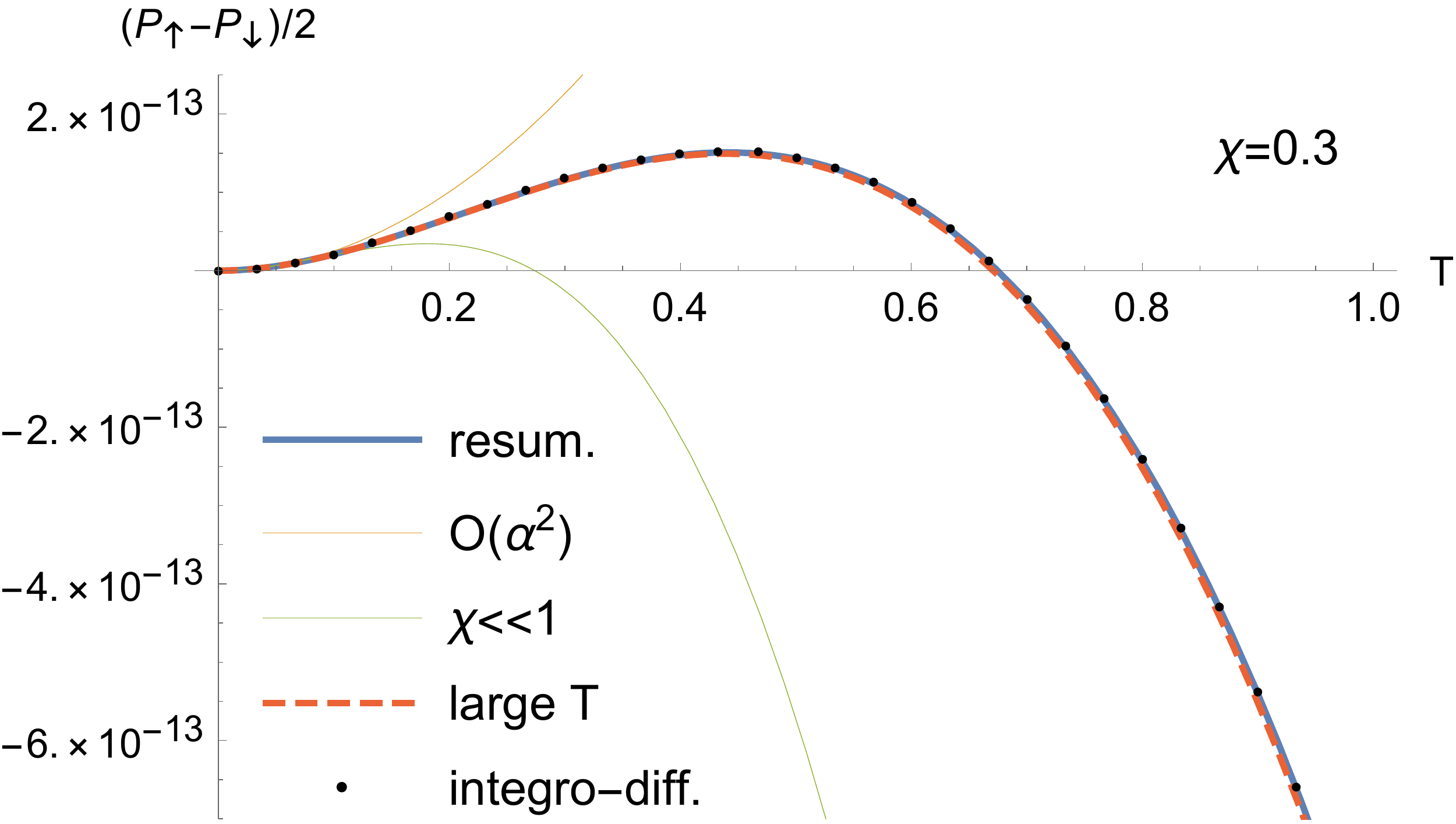}
\caption{Similar to Fig.~\ref{chi03plots} but for $(P_\downarrow-P_\uparrow)/2$, 
with Pad\'e approximant as in~\eqref{PadeHgen} $n=2$. The $\chi\ll1$ line gives the low energy limits in~\eqref{PdiffClassical}.}
\label{chi03plotDiff}
\end{figure}

\begin{figure}
\includegraphics[width=\linewidth]{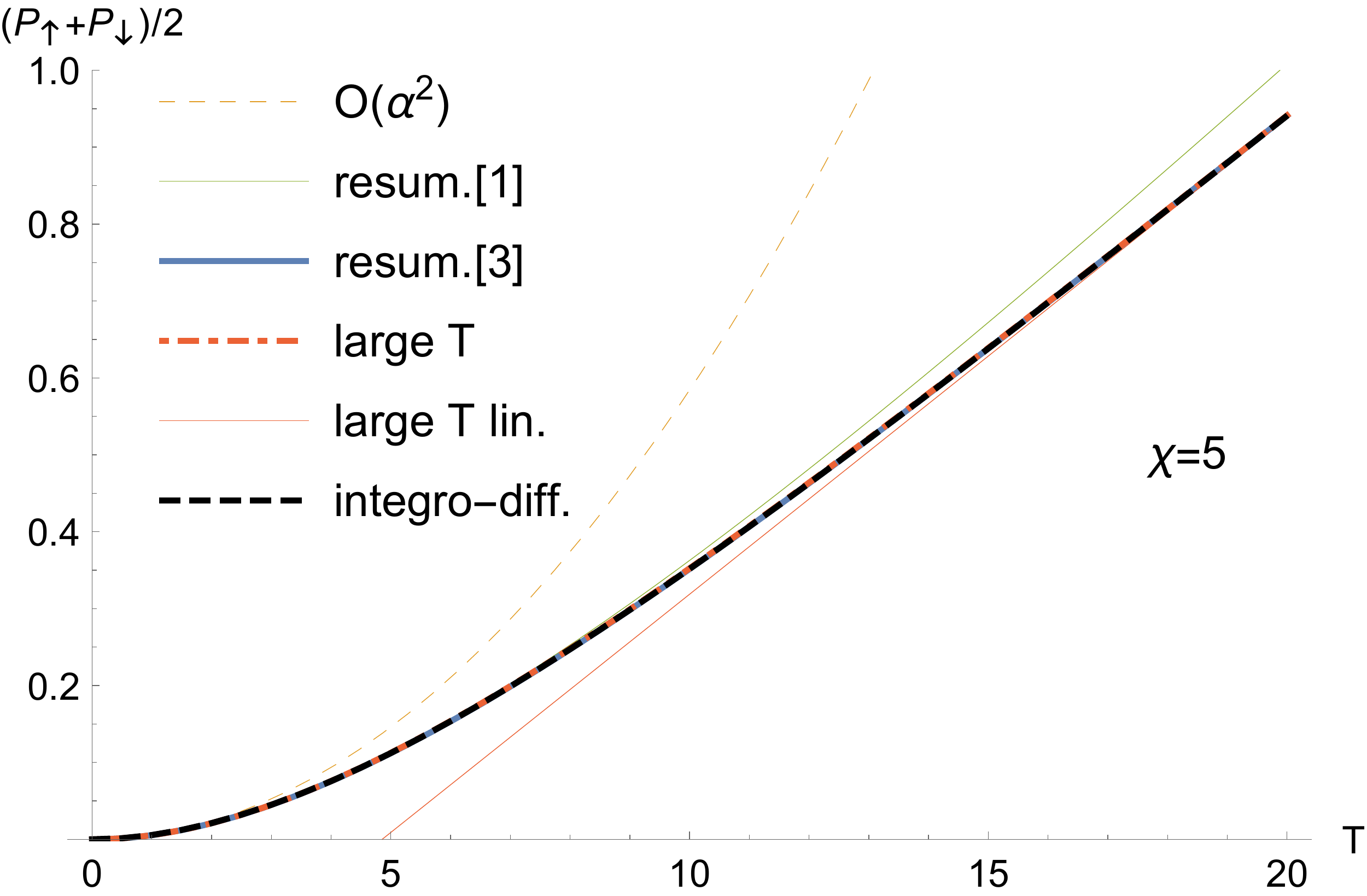}
\includegraphics[width=\linewidth]{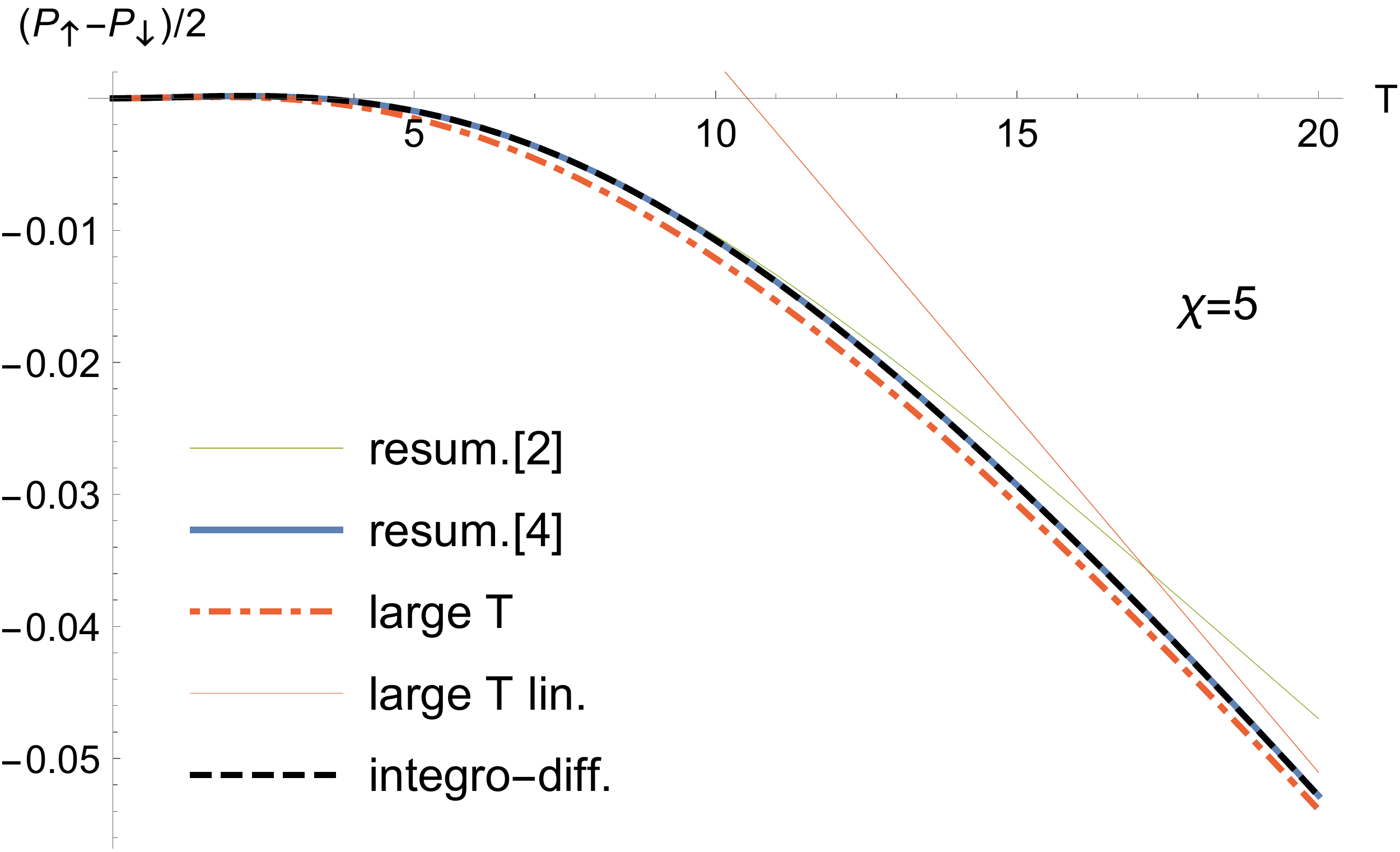}
\includegraphics[width=\linewidth]{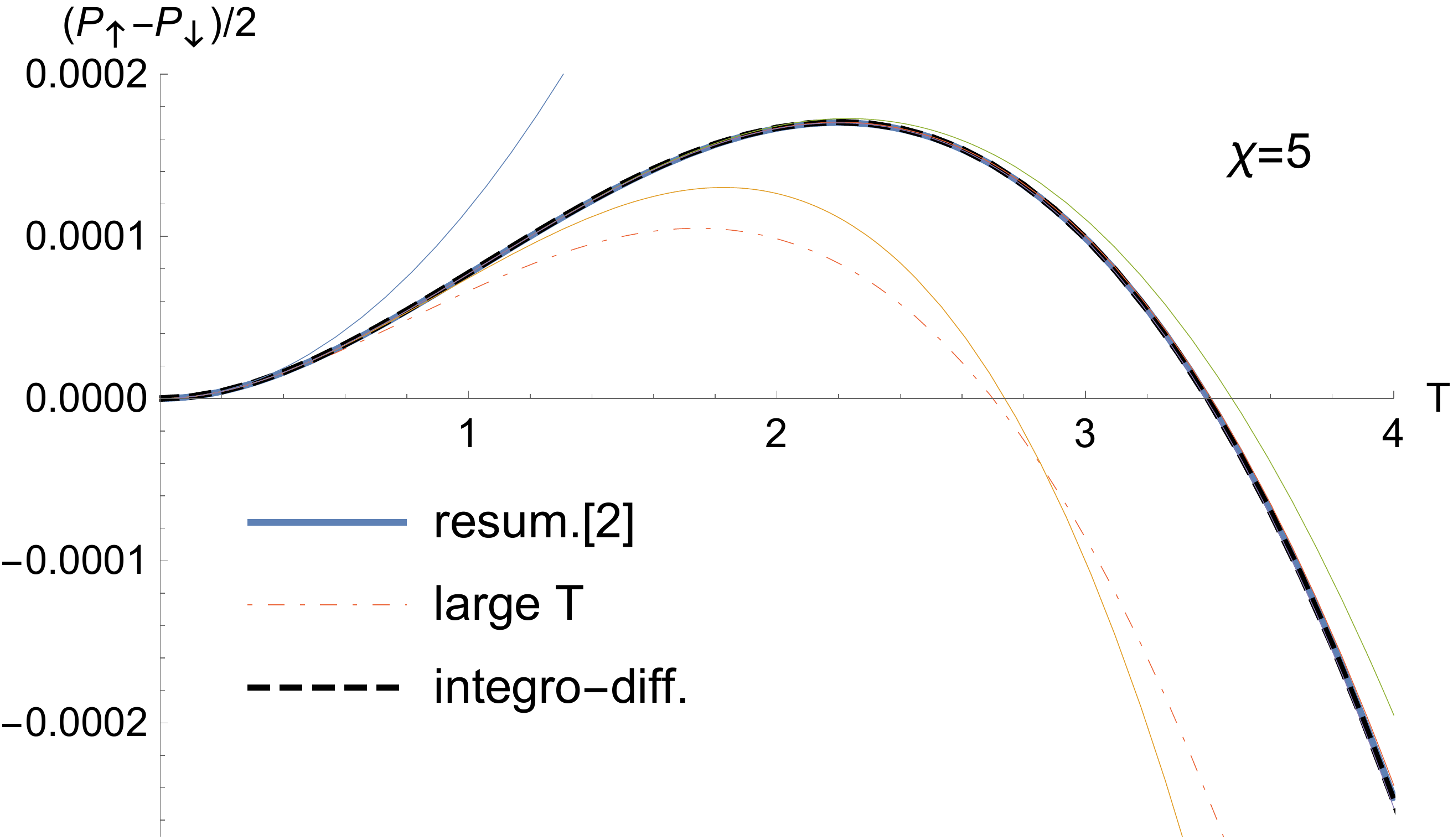}
\caption{As Fig.~\eqref{chi03plots} and~\ref{chi03plotDiff} but with $\chi=5$. $n$ in ``resum.$[n]$'' refers to the Pad\'e order in~\eqref{PadeHgen}. The thin solid lines in the last plot show the result of a direct summation (i.e. with no resummation) of the $\alpha$ expansion; summing more than the first $4$ or $5$ terms gives lines that agree, on the scale of that plot, with the results from resummation or from the integrodifferential equation.}
\label{chi5plots}
\end{figure}

\begin{figure}
\includegraphics[width=\linewidth]{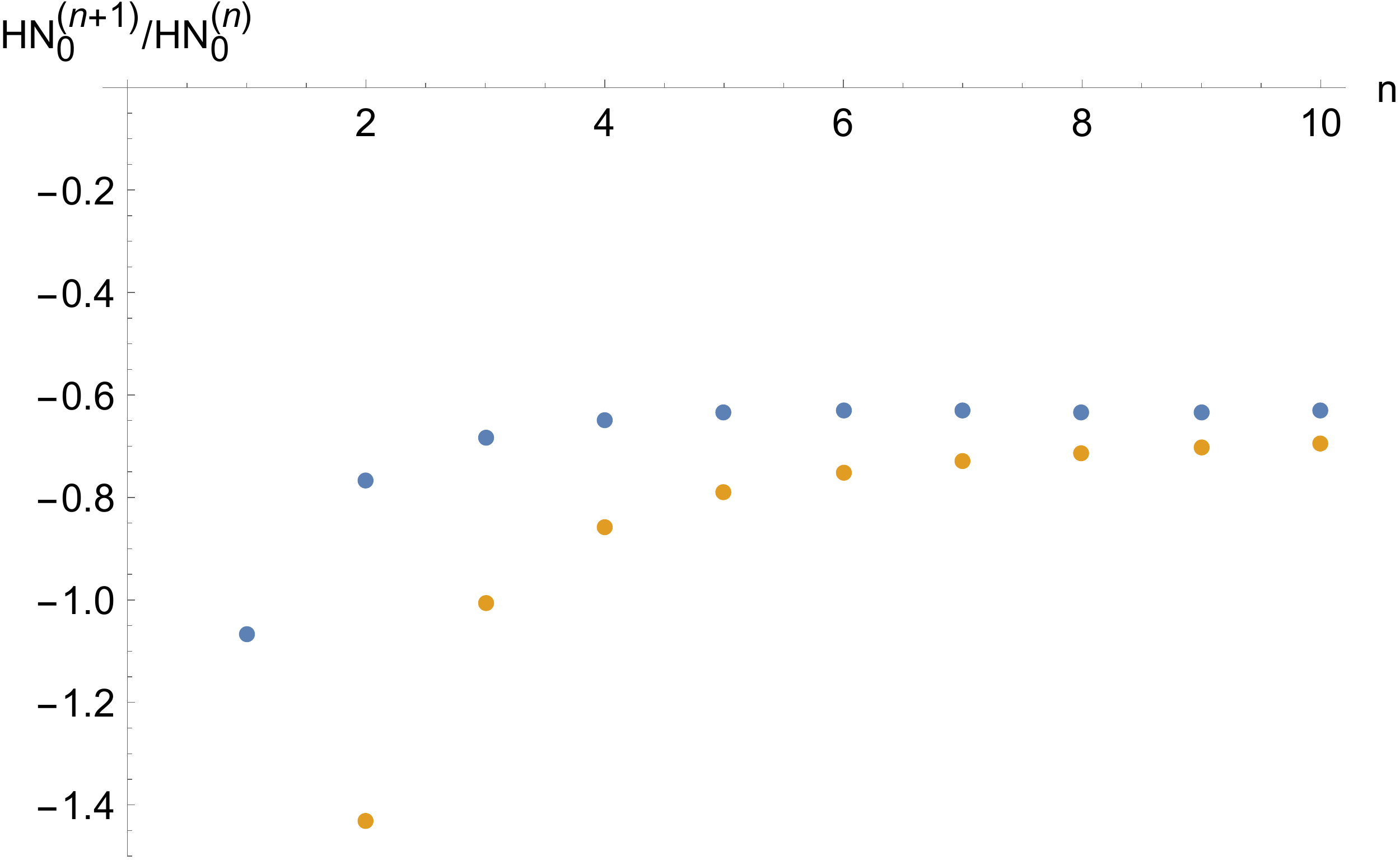}
\caption{Ratios of neighboring coefficients of $HN_0^{(n)}:=n!\{1,0\}\cdot{\bf N}^{(n)}$ for a Sauter pulse and $\chi=0.3$.}
\label{HNratiosFig}
\end{figure}

The $\chi$ expansions discussed above are asymptotic and can be resummed with the Borel-Pad\'e method~\cite{Costin:2019xql,Costin:2020hwg,Caliceti:2007ra,Florio:2019hzn,Dunne:2021acr,Baker1961,BenderOrszag,KleinertPhi4,ZinnJustinBook,Guillou1980}. There are other methods that can be more efficient~\cite{Mera:2018qte,Alvarez:2017sza,Torgrimsson:2020mto}, i.e. which require fewer terms to reach convergence. However, here we can without problem obtain a large number of terms in the $\chi$ expansions, so the standard Borel-Pad\'e method is enough. We will give a short summary of this method here. Another reason for doing so is to compare and contrast with the resummation method of convergent series discussed in Sec.~\ref{Resumming convergent series}. 

An asymptotic series is given by
\be\label{asymptoticSeries}
\psi(x)=\sum_{n=0}^\infty c_n x^n \;,
\ee
where $|c_n|\sim n!$ at large $n$. To resum this one can insert 
\be\label{LaplaceIntegral}
1=\frac{1}{n!}\int_0^\infty\ud t\; t^n e^{-t}
\ee
into the summand~\eqref{asymptoticSeries},
\be
\psi(x)=\int_0^\infty\!\ud t\, e^{-t}B\psi(xt) \;,
\ee
where
\be
B\psi(xt)=\sum_{n=0}^\infty\frac{c_n}{n!}(xt)^n
\ee
is the Borel transform. In the problems we are interested in we usually only have access to a finite number of terms, but $B\psi$ as a finite radius of convergence so the truncated transform
\be
B\psi_N(t)=\sum_{n=0}^N\frac{c_n}{n!}t^n
\ee
can be resummed by matching it onto a Pad\'e approximant,
\be
PB\psi(t)=\frac{\sum_{i=0}^I A_i t^i}{1+\sum_{j=1}^J B_jt^j} \;,
\ee
where the coefficients $A_i$ and $B_j$ are determined by demanding that
\be
PB\psi(t)=B\psi_N(t)+\mathcal{O}(t^{N+1}) \;.
\ee
One can choose different $I$ and $J$ depending on the problem, but $I=J$ or $I\approx J$ are often good choices. The final resummed result is then given by
\be
\psi(x)=\int_0^\infty\!\ud t\, e^{-t}PB\psi(xt) \;.
\ee

\section{Extra plots}

In Fig.~\ref{chi03plotDiff} we show the results for $(P_\downarrow-P_\uparrow)/2$ corresponding to the $(P_\downarrow+P_\uparrow)/2$ results in Fig.~\ref{chi03plots}. In Fig.~\ref{chi5plots} we show plots similar to Figs.~\ref{chi03plots} and~\ref{chi03plotDiff} but for $\chi=5$. One cannot actually neglect multiple pair production and other terms (multiple polarization/fermion loops) with similar exponential scalings for such a large $\chi$, which is obvious since the result for the probability is close to $1$. We present these results just to show the power of the resummation methods.
In Fig.~\ref{chi5plots} we can start to see a significant error at larger $T$ for $n=1,2$ in~\eqref{PadeHgen}. However, here we have increased $\chi$ so much that the results are no longer physical, and even then, the errors are not huge, so when we consider smaller $\chi$ (where we can neglect the fermion loops) the errors will be quite small. Thus, if we stick to a regime where our current approach gives physical results, then we find that we need very few terms from the $\alpha$ expansion to reach convergence. 

In Fig.~\ref{HNratiosFig} we plot the ratios of neighboring coefficients in the $\alpha$ expansion for the Sauter pulse case and with $\chi=0.3$. From this we can see that multiplying the coefficients by $n!$ seems to give a series with finite radius of convergence. 

\begin{figure*}
\includegraphics[width=\linewidth]{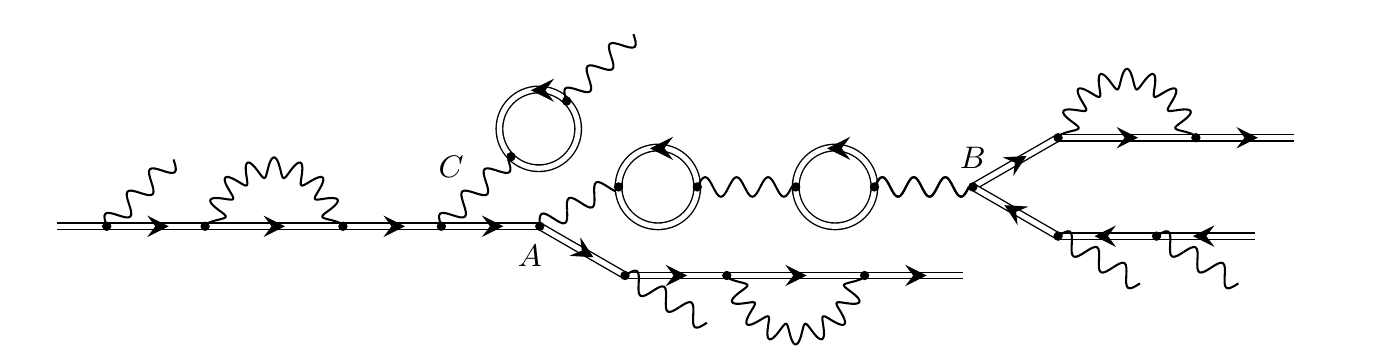}
\caption{Typical diagram for trident at higher orders.}
\label{diagramFig}
\end{figure*}

In Fig.~\ref{diagramFig} we give a diagramatic illustration of what processes and terms that are included and which ones are neglected. 
The particular diagram shown in Fig.~\ref{diagramFig} represents one typical process. We are interested in the infinite sum of the probabilities to produce one pair together with $0,1,2,3\dots$ photons. The amplitude to produce one pair and $n$ photons is itself given by an infinite coherent sum of amplitude terms with $0,1,2,3\dots$ loops. To leading order in $a_0$ or the pulse length, this coherent sum can be expressed as incoherent products of loop Mueller matrices~\cite{Torgrimsson:2020gws}. Thus, the sum over emitted photons and the sum over loops both lead to sums of incoherent products of Mueller matrices.
The photon emissions and loops on the electron line after vertex $A$, and on the fermion line connected to vertex $B$, do not contribute since we do not consider the momentum and spin of the final-state particles, which in this formalism follows from $({\bf M}^C+{\bf M}^L)\cdot\{1,0\}=\{0,0\}$. The sum over all fermion loops between vertex $A$ and $B$ can be expressed in compact form~\cite{Torgrimsson:2020gws}\footnote{See also~\cite{Bragin:2017yau,King:2016jnl,Meuren:2011hv} for different formulations of all-order birefringence~\cite{Bragin:2017yau,King:2016jnl} or quantities that correspond to sums over loops.}. One part of this loop sum can be neglected as long as production of more than one pair is negligible. The other part describes how the polarization of this intermediate photon changes as it propagates through the background field. This, vacuum birefringence part does not contribute here since we consider initial- and final-state fermions that are either unpolarized or polarized parallel (or antiparallel) to the magnetic field of the linearly polarized background. The sum of the fermion loops on externally emitted photons, e.g. the photon line starting at $C$, can again be neglected as long as multiple pair production is negligible. What is left is the sum over all photon emissions and loops attached to the electron line before vertex $C$.

\section{Large $T$ limit}\label{large T limit}

For a constant field, we can obtain the large $T$ results approximately by substituting the ansatz
\be\label{LNLeq}
{\bf N}(T,\chi)\approx T{\bf N}^L(\chi)+{\bf N}^{NL}(\chi) 
\ee
into~\eqref{integroDiffT}. The leading order is then determined by
\be\label{TLeq}
\begin{split}
\int_0^1&\frac{\ud q}{\chi}\left\{{\bf M}^C\cdot{\bf N}^L(\chi[1-q])+{\bf M}^L\cdot{\bf N}^L(\chi)\right\} \\
&=-2{\bf N}^{(2)}(\chi) \;,
\end{split}
\ee
and the next-to-leading order by
\be\label{TNLeq}
\begin{split}
\int_0^1&\frac{\ud q}{\chi}\left\{{\bf M}^C\cdot{\bf N}^{NL}(\chi[1-q])+{\bf M}^L\cdot{\bf N}^{NL}(\chi)\right\} \\
&={\bf N}^L(\chi) \;.
\end{split}
\ee
We solve these equations by expanding in $\chi$,
\be
{\bf N}^{L,NL}=\left\{\sum_{m=0}^{m_{\rm max}}A_m^{L,NL}\chi^m,\sum_{m=1}^{m_{\rm max}}B_m^{L,NL}\chi^m\right\}e^{-\frac{16}{3\chi}} \;,
\ee
where $A_m$ and $B_m$ are constants to be determined. Performing the $q$ integral in~\eqref{TLeq}, as explained above, gives a $\chi$ expansion that we then match with~\eqref{N2chiExpansion}. We find
\be
\begin{split}
&\{A_0^L,A_1^L,A_2^L,...\} \\
&=\{0.0439398,0.0839502,-0.157605,...\}
\end{split}
\ee  
and
\be
\begin{split}
&\{B_0^L,B_1^L,B_2^L,...\} \\
&=\{-0.0242686,0.0919387,-0.633781,...\} \;.
\end{split}
\ee
We have calculated terms up to $m_{\rm max}=25$. Next we solve~\eqref{TNLeq} in the same way and obtain $A_m^{NL}$ and $B_m^{NL}$. These coefficients grow factorially with alternating sign. We can therefore once again use Borel-Pad\'e to resum the $\chi$ expansions and obtain ${\bf N}^L(\chi)$ and ${\bf N}^{NL}(\chi)$.
The resulting approximation~\eqref{LNLeq} agrees well with the large $T$ limit of the exact result. 

\eqref{LNLeq} obviously breaks down for small $T$, since ${\bf N}\approx T^2{\bf N}^{(2)}$ for $T\ll1$. However, we can, without doing any extra calculations, significantly improve this approximation by simply making the replacement $aT+b\to F(T,+1), F(T,-1)$ or $(F(T,+1)+F(T,-1))/2$, where
\be\label{largeTandO2}
\begin{split}
F(T,\epsilon)&=aT+b\\
+&\left(\epsilon\sqrt{a^2+2bc}T-b\right)\exp\left(\frac{a+\epsilon\sqrt{a^2+2bc}}{b}T\right)
\end{split} \;.
\ee
In cases where the square root is complex, $(F(T,+1)+F(T,-1))/2$ is real.
The exponential term does not affect the results at large $T$ since it is exponentially suppressed compared to $aT+b$. But for $T\ll1$ we have, thanks to the added exponential, 
\be
F(T)\approx c T^2 \;,
\ee
so by choosing ${\bf c}$ such that ${\bf N}^{(\rm approx)}\approx T^2{\bf N}^{(2)}$ we have an approximation that is correct at both $T\gg1$ and $T\ll1$. Since ${\bf N}$ has a rather simple behavior, one can expect that ${\bf N}^{(\rm approx)}$ will not be far from ${\bf N}$ even at intermediate values of $T$. 
The results are shown in Fig.~\ref{chi03plots}, \ref{chi03plotDiff} and~\ref{chi5plots}. We can see that in all cases, using~\eqref{largeTandO2} indeed gives an improvement. From~\ref{chi03plotDiff} we see that for $\chi=0.3$ the improvement~\eqref{largeTandO2} gives a good precision even at intermediate values of $T$ where $(P_\downarrow-P_\uparrow)/2$ changes sign. However, from~\ref{chi5plots} we see that for $\chi=5$ the improvement~\eqref{largeTandO2} only gives a qualitative agreement at intermediate values of $T$. But this is not surprising since we should not expect to always be able to obtain a precise approximation using only the leading order in $T\ll1$ and the leading and next-to-leading orders in $T\gg1$. Since one anyway needs to include other diagrams (fermion loops) and processes to obtain physical results at such large values of $\chi$, and since we anyway can obtain good precision up to large $T$ by resumming only the $T\ll1$ expansion coefficients, we leave it to future studies to find ways to obtain higher orders in $T\gg1$ or to combine the leading and next-to-leading order in $T\gg1$ with more terms from the $T\ll1$ expansion.

\end{document}